\documentclass[a4paper]{jpconf}

\usepackage{graphicx}
\usepackage{epsf}
\usepackage{amsfonts}
\usepackage{amsmath}
\usepackage{bm}
\newcommand{\be}{\begin{equation}}\newcommand{\ee}{\end{equation}}
\newcommand{\bea}{\begin{eqnarray}}\newcommand{\eea}{\end{eqnarray}}
\newcommand{\brr}{\begin{array}}\newcommand{\err}{\end{array}}
\newcommand{\bit}{\begin{itemize}}\newcommand{\eit}{\end{itemize}}
\newcommand{\ben}{\begin{enumerate}}\newcommand{\een}{\end{enumerate}}

\newcommand{\bbm}{\begin{bmatrix}}\newcommand{\ebm}{\end{bmatrix}}
\newcommand{\ba}{\begin{array}}
\newcommand{\ea}{\end{array}}
\newcommand{\G}{\textbf}

\newtheorem{mydef}{Definition}
\newtheorem{Lemma}{Lemma}
\newtheorem{theorem}{Theorem}
\newcommand{\bd}{\begin{mydef}} \newcommand{\ed}{\end{mydef}}
\newcommand{\bthe}{\begin{theorem}} \newcommand{\ethe}{\end{theorem}}
\newcommand{\ble}{\begin{Lemma}} \newcommand{\ele}{\end{Lemma}}

\newcommand{\dr}{\mathrm{d}}

\def\ph{\varphi}
\def\lan{\langle}
\def\lf{\left}

\def\non{\nonumber}\def\pa{\partial}\def\ran{\rangle}

\def\ri{\right}
\def\al{\alpha}\def\ga{\gamma}
\def\de{\delta}\def\De{\Delta}

\def\la{\lambda}\def\La{\Lambda}\def\si{\sigma}\def\Si{\Sigma}
\def\om{\omega}

\def\1{{_{1}}}\def\2{{_{2}}}
\def\noHe0{:\;\!\!\;\!\!:H_e(0):\;\!\!\;\!\!:}
\def\noHm0{:\;\!\!\;\!\!:H_\mu(0):\;\!\!\;\!\!:}

\def\lan{\langle}
\def\lf{\left}

\def\non{\nonumber}
\def\pa{\partial}\def\ran{\rangle}

\def\ri{\right}

\def\al{\alpha}\def\ga{\gamma}
\def\de{\delta}\def\De{\Delta}

\def\la{\lambda}
\def\La{\Lambda}\def\si{\sigma}\def\Si{\Sigma}
\def\om{\omega}

\def\1{{_{1}}}\def\2{{_{2}}}

\begin{document}

\title{Classical space from quantum condensates}

\author{A Iorio $^{1}$ and L Smaldone$^{2}$}
\address{$^1$ Institute of Particle and Nuclear Physics, Faculty  of  Mathematics  and  Physics, Charles  University, V  Hole\v{s}ovi\v{c}k\'{a}ch  2, 18000  Praha  8,  Czech  Republic}
\address{$^2$ Institute of Theoretical Physics, Faculty of Physics, University of Warsaw ul. Pasteura 5, 02-093 Warsaw, Poland}

\ead{iorio@ipnp.troja.mff.cuni.cz}
\ead{Luca.Smaldone@fuw.edu.pl}

\begin{abstract}
We review the boson transformation method to deal with spontaneous symmetry breaking in quantum field theory, focussing on how it describes the emergence of extended and classical objects in such quantum context. We then apply the method to the emergence of space itself, as an extended and classical object resulting from the evaporation of a quantum black hole. In particular, we show how classical torsion and curvature tensors can emerge as effects of an inhomogeneous Nambu--Goldstone boson condensation in vacuum, in $E(3)$ invariant spinor models with symmetry breaking.
\end{abstract}

\section{Introduction}

It has been known for a long time that classical theory of elasticity, in the presence of underlying crystal defects, can be described as an Einstein--Cartan theory in 3-dimensions with its geometry-gravity tensors, metric, $g_{ij}$, curvature, $R^i_{jkl}$, and torsion, $T^i_{jk}$ \cite{kleinert1987,Kleinert:1989ky,Katanaev:1992kh,Zaanen:2021zqs}. Similar ideas are behind the description, by means of quantum field theory (QFT) in curved spacetime\footnote{For an approach based on deformed quantum algebras, see \cite{iorio2001quantization}.}, of the interaction between graphene quasi-particles of long-wavelength with the suitably engineered underlying carbon lattice, with disclination and dislocation topological defects \cite{IORIO20111334,Iorio:2011yz,Iorio:2013ifa,IorioReview,IorioPais,IORIO2018265,timeloopPRD2020,universe8090455}.
In principle, such emergent description could be derived from the underlying quantum electrodynamics, governing the elementary interactions between the electrons on the various molecular bands involved. In practice, this is not an easy task.

Within this general framework, it has been natural to put forward the proposal that both matter and space are emergent phenomena, stemming from the same fundamental quantum dynamics of basic constituents \cite{Acquaviva:2017xqi}. With Feynman \cite{feynman2011feynman} and Bekenstein \cite{bekenstein2003information}, we call such fundamental entities ``$X$ons'' \cite{Iorio_2019}. 

In Ref. \cite{Acquaviva:2020prd} this quantum gravity (QG) scenario, the ``QG quasiparticle picture'', was applied to the study of black-holes (BH) information paradox. In that approach, BH evaporation \cite{Hawking:1975vcx} is viewed a consequence of $X$ons interaction. The emergent effects are the space outside the BH, obtained at the expenses of the horizon area, and matter, seen as ``more traditional'' Hawking radiation. In other words, BH evaporating here is viewed as a ``factory'' where both space and matter are forged. In Ref. \cite{Acquaviva:2020prd} a connection between the fermionic nature of the $X$ons and the Bekenstein bound on entropy \cite{Bekenstein1981} was shown. 

Despite the encouraging results just recalled, a mechanism for the emergence of (classical) space and (quatum) matter, in this QG quasiparticle picture, was only given recently in Ref. \cite{spacefact}. There $X$ons dynamics is assumed to be given by a QFT model based on a fermion field, $\Psi(x)$, on $\mathbb{R}^4$, where the latter is just a target space, with no associated geometric structure. Therefore $x$ there is more similar to a label counting the $X$on's degrees of freedom than to a coordinate. In fact, the true space and its torsion and curvature, are seen to emerge after the spontaneous symmetry breaking (SSB) of the $E(3)$ symmetry of the field equation for $\Psi$. In that discussion, time is just an affine parameter, which does not play any active role.

The results of \cite{spacefact} rely heavily on the \emph{boson transformation method} (or simply \emph{boson method}) of Refs. \cite{Umezawa:1982nv,Umezawa:1993yq}. The core of that idea is that classical extended objects, such as solitons, surface boundaries or topological defects, can be described in QFT as macroscopic manifestations of an inhomogeneous boson condensation in the vacuum. In particular, topological defects in a system with SSB, are due to the condensation of the Nambu--Goldstone (NG) bosons. This idea was successfully applied to many areas of physics, including topological defects in crystals \cite{Wadati1977ASF,Wadati1978FDP,PhysRevB.18.4077}, solitons \cite{Blasone:2001aj}, vortices \cite{LepUme,LepManUme,Acquaviva:2020cjx}, and strings and bags in hadron physics \cite{TZE197563,PhysRevD.18.1192}.

The paper has the following structure: In Section \ref{genfram} we review the boson method. In Section \ref{emspace} we show that can be applied to describe emergent classical curvature and torsion. Finally, in Section \ref{conclusions} we propose our conclusions and perspectives.

\section{The boson method} \label{genfram}

To properly introduce the boson method, we first need to recall the definition and to give some examples of what is known as the \textit{Haag or dynamical map} \cite{Umezawa:1982nv, Umezawa:1993yq,Haag:1955ev,Greenberg:1994zu}.

Consider a real scalar field $\ph$ satisfying
\be
\lf(\Box+m^2\ri) \ph(x) \ = \ j[\ph](x) \, ,
\ee
where $j[\ph](x)$ is a functional of $\ph$.
Since in relativistic QFT the Fock space is defined in terms of asymptotic fields (e.g. $in$ fields), the dynamics is completely solved once the interacting operator $\ph$ is expanded in terms of asymptotic fields\footnote{Here and in the following we do not write explicitly field renormalization constants.}
\be \label{yfeqr}
\ph(x) \ = \  \ph_{in}(x) \ + \ \int \!\! \dr^4 y \, \Delta_R(x-y) \, j(y) \, ,
\ee
where  $\De_R(x-y)=[\ph(y) \, , \, \ph(x)] \, \Theta(x-y)$ is the retarded propagator. This relation, known as the \emph{Yang--Feldman equation} \cite{Yang:1950vi,greiner2013field}, only holds in a weak-sense \cite{greiner2013field}. The iterative solution of Eq.\eqref{yfeqr} is a series involving normal ordered products of the asymptotic fields, known as Haag or dynamical map
\cite{Umezawa:1982nv, Umezawa:1993yq,Haag:1955ev,Greenberg:1994zu}
\be
\ph(x) \ = \ v \ + \  \sum^\infty_{n=1} \, \int \!\! \dr^4 x_1 \, \ldots \, \dr^4 x_n \,  F_n(x;x_1,\ldots,x_n) \, : \ph_{in}(x_1) \, \ldots \, \ph_{in}(x_n): \, ,
\ee
$v$ being a constant. The coefficients of such expansion are truncated retarded Green's functions \cite{Umezawa:1982nv}:
\be
F_n(x;x_1,\ldots,x_n) \  \equiv \ (\Box_{x_1}+m^2) \ldots \,  (\Box_{x_n}+m^2) \, \lan R[ \ph(x) \, \ph(x_1) \, \ldots \, \ph(x_n)] \ran \, ,
\ee
where $\lan \ldots \ran \equiv \lan 0|\ldots |0\ran$.

In general one can consider a field $\Psi$ satisfying the equation of motion
\be \label{ineqp}
\La(\pa) \, \Psi(x) \ = \ j[\Psi](x) \,,
\ee
where $\La(\pa)$ is a differential operator, and rewrite it in the Yang--Feldman form
\be \label{yfeq}
\Psi(x) \ = \  \Psi_0(x) \ + \ \lf(\La^{-1} \star j[\Psi]\ri)(x) \, ,  \qquad \La(\pa) \Psi_0(x) \ = \ 0 \, ,
\ee
where $\La^{-1} \star j$ is the convolution of the interacting terms with a Green's function of the operator $\La$. The appropriate Green's function depends on the choice of $\Psi_0$. For example, in Eq.\eqref{yfeqr}, one has retarded Green's functions because $\Psi_0=\ph_{in}$. 

Although we gave a relativistic QFT example, the present considerations are far more general. In condensed matter physics, e.g. $\Psi_0$ is a quasiparticle field. Solving Eq.\eqref{yfeq}, we get the dynamical map, which is indicated as
\be \label{dmap}
\Psi(x) \ = \ \Psi[x;\Psi_0] \, .
\ee
The $\la \phi^4$ theory, with $\phi$ a complex scalar field,
\bea
{\cal L}(x)\, =\,  \pa_\mu  \phi^\dag(x) \pa^\mu\phi(x) - \mu^2
 \phi^\dag(x)\phi(x) - \frac{\la}{4} |\phi^\dag(x)\phi(x)|^2 \, ,
\eea
is considered in Ref. \cite{Blasone:1999aq}. As well known, this theory is invariant under the $U(1)$ phase transformation $\phi \to e^{i \al} \phi$, with $\al \in \mathbb{R}$ and SSB occurs when $\mu^2<  0$  and  $\lan  \phi  \ran  =  v=\sqrt{-2\mu^2/\la}$. 

Let us write the complex field in terms of two real fields, $\rho$ and $\chi$
\be
\phi(x) \ = \ (\rho(x)   +v) e^{i\chi(x)} \,,
\ee
whose equations of motion are
\bea
&&\lf[\Box -   (\pa_\mu \chi(x))^2  \,+\,m^2\ri]\rho(x) +  \frac{3}{2}m\,  g\,
\rho^2(x) + \frac{1}{2} g\,\rho^3(x)
\,=\, v \,(\pa_\mu \chi(x))^2 \, ,
\\ [2mm]
&&\pa_\mu \lf[ (\rho(x) + v )^2\pa^\mu \chi(x)\ri]\, =\, 0 \, ,
\eea
with $g=\sqrt{\la}$ and $m^2=\la v^2  > 0$. The field $\Psi_0$ is chosen to be the $in$-field $\phi_{in}$, satisfying the free field equation $\lf(\Box \, +\, m^2 \ri)\phi_{in}\, =\, 0 $ ($\La(\pa)=\Box \, +\, m^2$). Then, this is conveniently decomposed in a \emph{radial} part $\rho_{in}$ and in the phase $\chi_{in}$, the latter being the NG mode, as
\be \label{bjvortex}
\Psi_0(x)=\phi_{in}(x)\equiv (\rho_{in}(x)+v)\, e^{i\chi_{in}(x)}\, ,
\ee
satisfying the coupled equations
\bea
&&\lf[\Box - (\pa_\mu  \chi_{in}(x))^2 \,+\,m^2\ri](\rho_{in}(x) \, + \, v)
\,=\,  0 \, ,
\\ [2mm]
&&\pa_\mu \lf[(\rho_{in}(x) +v)^2 \,  \pa^\mu
\chi_{in}(x)\ri]\, =\, 0 \, .
\eea
The dynamical map can be written in the compact form
\bea \label{scalardm}
\phi(x) \, =\, T_C\lf\{(\rho_{in}(x) + v) e^{i\chi_{in}(x)} \exp\lf[-i\int_C
d^4y {\cal L}^I_{in}(y)\ri] \ri\} \, ,
\eea
where $T_C$ is the time-ordering along a Schwinger closed time-path from $t=-\infty$ to $t=+\infty$ and the way back, while ${\cal L}^I_{in}=(m^2-\mu^2)\phi^\dag_{in} \phi_{in}- \frac{\la}{4} |\phi_{in}^\dag\phi_{in}|^2=2m^2 v \rho_{in}-g m \rho^3_{in}-\frac{g^2}{4}\rho^4_{in}+\frac{5}{4}m^2 v^2$ is the (interaction picture) interacting Lagrangian \cite{Blasone:1999aq}.

In Section \ref{emspace} we will follow a similar approach, taking the decomposition
\be \label{psi0d}
\Psi_0(x) \ = \ \Psi_0[x;\psi(x),\ph_1(x), \ldots , \ph_N(x)]  \, ,
\ee
where $\psi$ (the `'radial mode") and $\ph_j$s (the NG modes) thus follow coupled field equations
\be \label{bjvortexeq2}
\La_\psi(\pa) \, \psi(x) \ = \ j_\psi[\psi,\ph_1, \ldots , \ph_N](x) \, , \qquad \La_{\ph_j}(\pa) \, \ph_j(x) \ = \ j_{\ph_j}[\psi,\ph_1, \ldots , \ph_N](x)  \, ,
\ee
$j=1,\ldots, N $.

We have now all the needed ingredients to introduce the boson method, that is the main goal of this Section. 
 
From Eq.\eqref{dmap} and the decomposition \eqref{psi0d}, we get
\be \label{dmapsi}
\Psi(x) \ = \ \Psi[x;\psi(x),\ph_1(x), \ldots , \ph_N(x)] \, .
\ee
The canonical transformation
\be
\ph_j(x) \ \to \ \ph_j^f(x) \ \equiv \ \ph_j(x) \ + \ f_j(x) \, ,
\ee
with $f_j$ being $c$-number functions, is known as \emph{boson transformation}, which describes the condensation of $\ph_j$ quanta in the vacuum. Then, the \emph{boson transformation theorem} \cite{Matsumoto:1979hs,Umezawa:1982nv} asserts that
\be
\Psi^f(x) \ = \ \Psi[x;\psi(x),\ph^f_1(x),\ldots,\ldots,\ph^f_{N}(x)] \, ,
\ee
when $\ph^f_j$ satisfies the same equation as $\ph_j$, is still a solution of the original dynamical problem, i.e.
\be
\La(\pa) \Psi^f(x) \ = \ j\lf[\Psi^f\ri](x) \, .
\ee
The solution $\Psi^f$ describes the creation of self-sustained and extended objects, as macroscopic manifestation of the condensation in the vacuum of a large number of $\ph_j$ quanta. Physically relevant cases are those for which the $f_j$s are not Fourier-transformable, i.e. when $\lim_{|x|\to \infty}f_j \to \infty$ or
\be \label{ctd}
\lf[\pa_\mu \, , \pa_\nu \ri] f_j(x) \ \neq \ 0 \, ,
\ee
at some point $x$.

The condition \eqref{ctd} can only be fulfilled when the $\ph_j$s are gapless fields, as in the case of NG bosons \cite{Umezawa:1982nv, Umezawa:1993yq}.
Therefore, we turn our attention to field equations that have a symmetry group $G$, with Lie algebra
\be \label{lieal}
\lf[t_a \, , \, t_b \ri] \ = \ i C_{a b c} \, t^c \,, \qquad a,b,c = 1, ..., {\rm ord} \, G \, .
\ee
A system undergoes SSB when
\be \label{qonv}
Q_a |0\ran \ \neq \ 0 \, ,
\ee
for some $a$s, i.e., when some of the Noether charges do not annihilate the vacuum. We will then perform the boson transformation of the corresponding NG bosons
\be
\ph^a(x) \ \to \ \ph^a(x) \ + \ f^a(x) \, ,
\ee
into the dynamical map \eqref{dmapsi}. Note that we now use the group index to properly count the NG fields.

When $\Psi \to \Psi^f$, one has the decomposition \cite{Umezawa:1982nv,Matsumoto:1980uu}
\be
 \Psi^f(x) \ = \  U(x) \, \tilde{\Psi}(x)  \, ,
\ee
where $U(x) \equiv U[f(x)]$ is a local gauge transformation, whose rigid generator belongs to $G$, and where either $\tilde{\Psi}$ does not contain $f$ or it only contains $f$ through regular expressions (usually through regular derivatives, $\pa f$).
It is also useful to introduce \cite{Umezawa:1982nv,Matsumoto:1980uu}
\bea
A_\mu(x)       & = & i \, U^{-1} (x) \pa_\mu U (x) \,, \label{am} \\[2mm]
F_{\mu \nu}(x) & = & i \, U^{-1} (x) \lf[\pa_\mu \, , \pa_\nu \ri] U(x) \,, \label{fmn}
\eea
giving
\be \label{variousf}
F_{\mu \nu}(x) \ = \ \pa_\mu A_{\nu}(x)-\pa_\nu A_{\mu}(x) \, - \, i \, \lf[A_\mu(x) \, , \, A_\nu(x)\ri] \, .
\ee
If we write $U$ in the form
\be \label{uform}
U(x) \ = \ \exp\lf(-i \, \al^a(x) \, t_a\ri)  \, ,
\ee
where $\al^a(x)$ are functionals of $f^a$, then
\bea
A_\mu{}^a(x) & = &  \pa_\mu \al^a(x) \, , \\[2mm]
F_{\mu \nu}{}^a(x) & = & \pa_\mu A_{\nu}{}^a(x)-\pa_\nu A_{\mu}{}^a(x) \, +  \, C_{a b c} \, A_\mu{}^{b} (x) \, A_\nu{}^c (x) \, , \non \\[2mm]
& = & \lf[\pa_\mu \, , \, \pa_\nu \ri] \al^a(x) \, +  \, C_{a b c} \,\pa_\mu \al^b(x) \, \pa_\nu \al^c(x) \, , \\[2mm]
A_\mu(x) & = & A_\mu{}^a(x) \, t_a \, , \qquad F_{\mu \nu}(x) \ = \ F_{\mu \nu}{}^{a}(x) \, t_a \, .
\eea
Let us remark that $F_{\mu \nu}{}^a(x)$ has a term arising from the nonabelian structure of the Lie group manifold $G$.

$A_\mu{}^a$ is a classical gauge field which is a physical manifestation of the inhomogeneous condensate of $\ph^a$ quanta. Since $F_{\mu \nu}{}^a$ is physically observable, one requires regularity:
\be \label{bide}
\lf[\pa_\mu \, , \, \pa_\nu\ri] A_\la{}^a(x) \ = \ \lf[\pa_\mu \, , \, \pa_\nu\ri] \pa_\la \al^a (x) \ = \ 0 \, , \qquad \lf[\pa_\mu \, , \, \pa_\nu\ri] F_{\la \rho}{}^a(x) \ = \ 0    \, .
\ee

Before moving to the application of the boson method to the QG scenario of our interest here, let us go back to the example of the $\la \phi^4$ theory. In that case we have only one NG field $\chi_{in}$ (see Eq.\eqref{bjvortex}). Therefore, we perform the boson transformation
\be
\chi^f_{in}(x) \ = \ \chi_{in}(x) \,-\,f(x) \, ,
\ee
in the dynamical map \eqref{scalardm}:
\bea
\psi^{f}(x)     \,=\, e^{-i f(x)}
T_C\lf\{(\rho_{in}(x)+v)e^{i\chi_{in}(x)}
\exp\lf[-i\int_C d^4y {\cal L}^{I}_{in}(y)\ri] \ri\} \,.
\eea
Hence, in this case
\be
U(x) \ = \ e^{-i f(x)} \, , \qquad A_\mu(x) \ = \ \pa_\mu f(x) \, , \qquad F_{\mu \nu}(x) \ = \ [\pa_\mu,\pa_\nu]f(x) \, .
\ee
A relativistic string solution along the $x_3$-axis is obtained when $f$ is the azimuthal angle in cylindrical coordinates, $f=\theta$. In that case, $ \nabla \times \nabla \theta \ = \ 2 \pi \hat{e}_3 \, \de(x_1)\de(x_2)
$, where $\hat{e}_3$ is the unit vector along the $x_3$-axis, hence $A_\mu=\pa_\mu \theta$ and $F_{12}(x)=2\pi \de(x_1) \de(x_2)$, the other components being zero.

\section{Classical geometry from $E(3)$ symmetry breaking} \label{emspace}

Our goal now is to apply the boson transformation method to the case of a QFT of a fermion, $\Psi(x)$, $ x = (\si,\G x) = (\si,x^1,x^2,x^3) \in \mathbb{R} \times \mathbb{R}^3$. As extensively explained in Ref.\cite{spacefact}, this is a good description of the $X$ons model of Refs. \cite{Acquaviva:2017xqi,Iorio_2019,Acquaviva:2020prd,universe8090455}, for an observer whose energy scale is ``low'', compared to Planck's. The presentation will closely follow the one of Ref.\cite{spacefact}.

Let us take the $X$on field, $\Psi(x)$, in the irreducible spinor representation of $SO(3)$
\be
\Psi(x) \ = \ \begin{pmatrix} \Psi_\uparrow(x) \\ \Psi_\downarrow(x) \end{pmatrix} \,,
\ee
and let us assume that the field equation \eqref{ineqp} for $\Psi$ is invariant under $E(3)$. The $e(3)$ Lie algebra is
\bea \label{po1}
&& \lf[P_a \, , \, P_b \ri] \ = \ 0 \, ,\\[2mm] \label{po2}
&& \lf[P_a \, , \, J_{b}\ri] \ = \ i \, \varepsilon_{a b c} P_c  \, , \\[2mm] \label{po3}
&& \lf[J_{a} \, , \, J_{b}\ri] \ = \ i \, \varepsilon_{a b c} \, J_{c} \,,
\eea
where $a,b,c=1,2,3$.

When we perform rotations on $\Psi(x)$ we have
\bea
\Psi'(x) \ = \ e^{i \, \theta^a \, J_a} \, \Psi(x) \, e^{-i \, \theta^a \, J_a} \ = \  e^{-\frac{i}{2} \theta^a \si_a} \, \Psi(\si,\G x') \, , \qquad x'^a \ \equiv \ R^a{}_b(\theta) x^b \, .
\eea
Here the $\theta^a$s are real parameters, $R$ belongs to the fundamental representation of $SO(3)$ and $\si_a$ are the Pauli matrices. 

When we perform translations on $\Psi(x)$ we have
\be
\Psi'(x) \ = \ e^{i \, u^a \, P_a} \Psi(x) e^{-i \, u^a \, P_a} \ = \ \Psi(\si,\G x') \, , \qquad x'^a \ \equiv \ x^a +u^a \, ,
\ee
where the $u^a$s are real parameters.

The relevant issue here is to understand how to count NG modes \cite{Brauner:2010wm,Watanabe:2013iia}.
In fact, it is only in the direction where the symmetry is broken that one can produce a nontrivial (classical) gauge field $A_i{}^a$ . A general result was presented in Ref. \cite{Watanabe:2013iia}. There, it was proved that a weak linear dependence condition of the form
\be  \label{redcon}
\int \!\! \dr^3 x \, \sum_a \, c_a(x) \, j_a^0(x) |0\ran \ = \ 0 \, ,
\ee
implies a redundancy in the number of NG bosons, which are reduced with respect to the number of broken generators (see Eq.\eqref{qonv}). In Eq.\eqref{redcon} $j_a^0$ are the charge densities corresponding to the broken generators $Q_a$ (see Eq.\eqref{qonv}), and $c_a$ are generic functions.

Let us write $P_a=\int \!\! \dr^3 x p_a(x)$, $J_a=\int \!\! \dr^3 x j_a(x)$. The total angular momentum $J_a$ can be decomposed into the orbital and the spin part $J_a=L_a+S_a$, with $L_a= \int \!\! \dr^3 x l_a(x)$, $S_a \ = \ \int \!\! \dr^3 x \Si_a(x)$. 

For a scalar field, $J_a=L_a$. 
Since $l_a(x) = \varepsilon_{a b c} x^b p_c(x)$, Eq.\eqref{redcon} is thus fulfilled in a strong sense and only $6 - 3 = 3$ independent NG bosons are present. 
Eq.\eqref{redcon} is also fulfilled when $S_a|0\ran=0$ and only three NG modes are present. 

Here the key observation is that, when $S_a|0\ran \ \neq \ 0$,
\be \label{ourcasej}
\int \!\! \dr^3 \, x \, j_a(x)|0\ran \ = \ \int \!\! \dr^3 x \, \varepsilon_{a b c} x^b p_c(x)|0\ran \, + \, \int \!\! \dr^3 x \, \Si_a(x)|0\ran \, .
\ee
$\Si^a$ and $p^a$ are linearly independent, thus $j_a$ and $p_a$ are, in general, linearly independent. In this case six NG bosons should appear in the spectrum. 

In the following we will briefly review the two cases reported in the Table.

\begin{table}[h]
\centering
\label{tabu1}
\caption{Main features of the different SSB phases.}
\begin{tabular}{|c|c|c|}
\hline \hline
 $$ & $ S|0\ran \ = \ 0 $ & $ S|0\ran \ \neq \ 0$\\
\hline
NG fields & $X^a$ & $X^a$ \, , \, $\Theta^a$ \\[2mm]
\hline
 $A_i^a$ &  $ e_i{}^a \ \neq \ \de_i^a \, , \quad \om_i{}^a \ = \ 0 $  & $ e_i{}^a \ \neq \ \de_i^a \, , \quad \om_i{}^a \ \neq \ 0 $   \\
\hline
 $F_{i j}{}^a$ &  $T_{ij}{}^a \ \neq \ 0 \, , \quad R_{il}{}^a \ = \ 0$  & $T_{ij}{}^a \ \neq \ 0 \, , \quad R_{il}{}^a \ \neq \ 0$\\
\hline
\end{tabular}
\end{table}

%
\subsection{The case $P_a|0\ran \neq 0$, $S_a |0\ran = 0$} \label{trancase}

Let us first consider
\be
P_a \, |0\ran \ \neq \ 0 \, , \qquad S_a \, |0\ran \ = \ 0 \, .
\ee
The dynamical map can be given in the compact form of Eq.\eqref{dmap} and we decompose $\Psi_0(x)$ as
\be \label{eqpsi0}
\Psi_0(x) \ = \ \Psi_0[x; \psi(x), X^a(x)] \, .
\ee
$X^a(x)$ ($a=1,2,3$) are the three NG modes which are the analogues of acoustic phonons in crystals \cite{Wadati1977ASF,Wadati1978FDP,PhysRevB.18.4077, Watanabe:2013iia} (\emph{space-phonons}), while our minimal assumption only allows for one type of fermionic matter $\psi(x)$, that is our radial mode.

Since $\La(\pa) \Psi_0(x)=0$, $\psi$ and $X^a$ obey coupled equations (see Eq.\eqref{bjvortexeq2}):
\be \label{coeq}
\La_\psi(\pa) \psi(x) \ = \ j_\psi[\psi,X^a](x) \, , \qquad \La^a_X(\pa) X^a(x) \ = \ j_X[\psi,X^a](x) \, .
\ee

The boson transformation is then\footnote{Here we consider only static transformations, i.e. $\si$ does not play any active role.}
\be \label{xbostr}
X^a(x) \ \to \ X_u^a(\G x) \ \equiv \ X^a(x) \, + \, u^a(\G x) \, ,
\ee
where $X_u^a(x)$ satisfy the same equations as $X^a(x)$. By definition
\be\label{Psiu}
\Psi^u(x)  \ = \  \Psi[x; \psi(x),X^a_u(x)] \, ,
\ee
which satisfies the field equation $\La(\pa)\Psi^u=0$, because of the boson transformation theorem. Moreover, (\ref{Psiu}) can be written as $\Psi^u=U \tilde{\Psi}$, with $[\pa_i, \pa_j] \tilde{\Psi}=0$, and
\be \label{utransl}
U(\G x) \ = \ \exp\lf(- i \, y^a(\G x) \, \mathcal{P}_a\ri) \, .
\ee
$\al^a(\G x)=y^a(\G x)$ is a functional of $u$ so that, in general, $[\pa_i, \pa_j] y^a(\G x) \neq 0$ and
where $\mathcal{P}_a$ are a $2 \times 2$ matrix representation of the translations algebra. Note that we now indicate the indices of the coordinate with $i,j,k$. The reason is clear when looking at the classical gauge potential
\be
A_j(\G x) \ = \ e_j{}^{a}(\G x) \, \mathcal{P}_a \, ,
\ee
with
\be
e_j{}^{a}(\G x) \ \equiv \ \pa_j y^{a}(\G x) \, .
\ee
Then $y^a$ should be viewed as a set of flat coordinates (the metric $\de_{a b}$ permits to contract group indices) and then $e_j{}^{a}(\G x)$ are identified with \emph{Vielbeins}. The regularity condition \eqref{bide} now reads
\be
\lf[\pa_i \, , \, \pa_j\ri]e_k{}^a(\G x) \ = \ \lf[\pa_i \, , \, \pa_j\ri]\pa_k y^a(\G x) \ = \ 0 \, .
\ee
As a consequence, a metric tensor can be naturally defined on the configuration space
\be \label{gdef}
g_{i j}(\G x) \ \equiv \  \delta_{a b} \, \pa_i y^{a}(\G x) \, \pa_j y^{b}(\G x) \ = \ \delta_{a b} \, e_i{}^{a}(\G x) \, e_j{}^{b}(\G x) \, .
\ee

Finally, one can compute the field strength Eq.\eqref{variousf}
\bea
F_{i j}(\G x) \ = \ T_{i j}{}^a(\G x) \, \mathcal{P}_a \, ,
\eea
where
\be
T_{i j}{}^{a}(\G x) \ = \ \pa_i e_j{}^{a}(\G x)-\pa_j e_i{}^{a}(\G x) \, ,
\ee
is the classical torsion tensor. We have then shown that \emph{classical torsion emerges as a ``macroscopic'' manifestation of the condensation of the space-phonons}.

\subsection{The case $P_a|0\ran \neq 0$, $S_a |0\ran \neq 0$} \label{comcase}

When
\be \label{comcond}
P_a \, |0\ran \ \neq \ 0 \, , \qquad S_a \, |0\ran \ \neq \ 0 \, ,
\ee
we are spontaneously breaking the whole $E(3)$ symmetry, therefore six NG fields $X^a(x),\Theta^a(x)$ appear in the physical spectrum. Let us just stress once more that the rotational NG field is only possible because of the spin structure.

The dynamical map for the $X$on field is
\be \label{eqpsi02}
\Psi(x) \ = \ \Psi[x; \psi(x), X^a(x), \Theta^a(x)] \, .
\ee
The boson transformations
\be \label{xtbtran}
X^a(x) \ \to \ X^a_u \ \equiv  \ X^a(x) \, + \, u^a(\G x) \, , \qquad \Theta^a(x) \ \to \ \Theta_\theta^a(\G x) \ \equiv \  \Theta^a(x) \ + \ \theta^a(\G x) \, ,
\ee
lead to
\be
\Psi^{u,\theta}(x) \ = \  \Psi_0^{u,\theta}(x) \ + \ \ldots \, , \qquad
\Psi^{u,\theta}_0(x) \ = \ \Psi_0[x; \psi(x), X^a_u(x), \Theta^a_\theta(x)] \, ,
\ee
at the linear order of the dynamical map.

The gauge field $U(x)$ can be written as
\be \label{utransl1}
U(\G x) \ = \ \exp\lf(-i \, \theta^{a}(\G x) \, \mathcal{J}_a-i \, y^a(\G x) \, \mathcal{P}_a\ri) \, ,
\ee
with $\mathcal{P}_a$ and $\mathcal{J}_a$ forming a $2 \times 2$ matrix representation of the $e(3)$ Lie algebra \eqref{po1}-\eqref{po3}.
Therefore
\be
A_j(\G x) \ = \ \om_j{}^{a}(\G x) \, \mathcal{J}_a \ + \ e_j{}^{a}(\G x) \, \mathcal{P}_a \, .
\ee
It is clear that $\om_j{}^a$ should be identified with the \emph{spin connection}, hence the field strength \eqref{variousf} is
\bea
F_{i j}(\G x) \ = \ R_{i j}{}^{a}(\G x) \, \mathcal{J}_a \, + \, T_{i j}{}^a(\G x) \, \mathcal{P}_a \,,
\eea
where the curvature and the torsion tensors have the usual expression
\bea
R_{i j}^{a}(\G x) & = & \pa_i \om_j{}^{a}(\G x)-\pa_j \om_i{}^{a}(\G x) \ + \ \varepsilon_{a b c} \,  \om_i{}^{b} (\G x) \om_j{}^{c}(\G x) \, , \\[2mm]
T_{i j}{}^{a}(\G x) & = & \pa_i e_j{}^{a}(\G x)-\pa_j e_i{}^{a}(\G x) \ + \ \varepsilon_{a b c} \, e_i{}^{b}(\G x) \, \om_j{}^{c}{}(\G x)  \,.
\eea
We have then proved that \emph{a low-energy observer experiences non-trivial curvature and torsion, due to the non-trivial boson condensation of NG fields after the full SSB of $E(3)$}. The detailed dynamics clearly depends on the interaction with the matter $\psi$.

In the case of small $u^a$, $y^a(\G x) \approx x^a+u^a(\G x)$ and it is reasonable to use the decomposition \cite{spacefact}
\be
\Psi_0(x) \ = \ e^{-i \lf[\lf(x^a+X^a(x) \ri)\mathcal{P}_a+\Theta_a(x) \mathcal{J}_a\ri]} \, \psi(x) \,,
\ee
that, after the boson transformation, gives
\be
\Psi^{u,\theta}_0(x) \ = \ e^{-i \lf[\lf(y^a(\G x)+X^a(x)\ri) \mathcal{P}_a+\lf(\Theta_a(x)+\theta^a(\G x)\ri) \mathcal{J}_a\ri]} \, \psi(x) \, .
\ee
If $\La(\pa) = \ga^\mu \, \pa_\mu$, $\ga^\mu$ being $2 \times 2$ matrices, an average of the field equation $\La(\pa) \Psi^{u,\theta}_0(x)=\ga^\mu \, \pa_\mu \Psi_0^{u,\theta}(x) \ \ =\ 0$ with respect to the NG modes gives
\begin{flushright}

\end{flushright}
\be \label{psicurved}
i \, \ga^\mu D_\mu \, \psi(x) \ = \ 0 \, , \qquad D_\mu \ = \ \lf(\pa_\si, \pa_j+e_j{}^a(\G x) \, \mathcal{P}_a+\om_j{}^a(\G x) \, \mathcal{J}_a \ri) \, ,
\ee
which is the equation of a massless fermion field coupled with Vielbeins and spin connection \cite{IORIO2018265}.
\section{Conclusions} \label{conclusions}

In this paper we reviewed the fundamental notion of dynamical map and the boson transformation method. Then we showed how this machinery can be employed to explain the emergence of classical geometric tensors as a low-energy manifestation of a non-trivial condensation in vacuum of the NG bosons in a $E(3)$-symmetric model with SSB. When translation symmetry is broken, these are the analogues of acoustic phonons in a crystal, while a non-trivial spin structure is necessary in order to have three independent NG fields related to the breaking of rotation symmetry. In the view of Ref.\cite{spacefact}, this mechanism is realized in the QG quasiparticle picture (see Refs. \cite{Acquaviva:2017xqi,Acquaviva:2020prd}), but it can be equally well-applied in other scenarios as condensed matter analogs of QG \cite{IORIO20111334,Iorio:2011yz,Iorio:2013ifa,IorioReview,IorioPais,IORIO2018265,timeloopPRD2020,universe8090455}.

An apparently unrelated issue is the the description of oscillations of neutrinos in QFT. In that context, a condensate structure of vacuum, which breaks Poincar\'e invariance, naturally appears \cite{ALFINITO199591,Blasone:2019rxl,bigs2,Smaldone:2021mii}, and a dynamical origin of such condensate from an underlying fermionic dynamics with SSB has been extensively investigated \cite{Mavromatos:2009rf,Mavromatos:2012us,bigs1}. It is then worth to investigate the connection with $X$ons.

\section*{Acknowledgements}
\vspace{2mm}

A.I. acknowledges support from Charles University Research Center (UNCE/SCI/013). L.S. was supported by the Polish National Science Center grant 2018/31/D/ST2/02048 and thanks the Institute of Particle and Nuclear Physics of Charles University for the kind hospitality while parts of this work were completed.
\section*{References}
\vspace{2mm}


\bibliography{librarySvN}{}
\bibliographystyle{iopart-num}
\end{document}